\documentclass[aps,preprint]{revtex4}%
\usepackage{amsfonts}
\usepackage{amsmath}
\usepackage{amssymb}
\usepackage{graphicx}%
\newtheorem{theorem}{Theorem}
\newtheorem{acknowledgement}[theorem]{Acknowledgement}

\begin{document}
\preprint{ }
\title[ ]{A Configurationally-Resolved-Super-Transition-Arrays method for calculation of
the spectral absorption coefficient in hot plasmas.}
\author{G. Hazak}
\affiliation{Physics Department, Nuclear Research Center - Negev P.O. Box 9001, Beer -
Sheva, ISRAEL}
\author{Y. Kurzweill}
\affiliation{Physics Department, Nuclear Research Center - Negev P.O. Box 9001, Beer -
Sheva, ISRAEL}

\begin{abstract}
A new method, 'Configurationally-Resolved-Super-Transition-Arrays', for
calculation of the spectral absorption coefficient in hot plasmas is
presented. In the new method, the spectrum of each Super-Transition-Array is
evaluated as the Fourier transform of a single Complex Pseudo Partition
Function, which represents the \textit{exact} analytical sum of the
contributions of all constituting unresolved transition arrays sharing the
same set of one-electron solutions. Thus, in the new method, the spectrum of
each Super-Transition-Array is resolved down to the level of the (unresolved)
transition arrays. It is shown that the corresponding spectrum, evaluated by
the traditional Super-Transition-Arrays (STA) method [A. Bar Shalom, J. Oreg,
W.H. Goldstein, D. Shvarts and A. Zigler, Phys. Rev. A \textbf{40}, 3183
(1989)], is just the coarse grained Gaussian approximation of the
Configurationally-Resolved-Super-Transition-Array. A new computer program is
presented, capable of evaluating the absorption coefficient by both the new
configurationally resolved and the traditional \ Gaussian
Super-Transition-Arrays methods. A numerical example of \ gold at temperature
$1keV$ and density $0.5$~$gr/cm^{3}$ , is presented, demonstrating the
simplicity, efficiency and accuracy of the new method.

\end{abstract}
\date{\today}
\startpage{1}
\endpage{102}
\maketitle

\section{Introduction}

The radiative opacity is an essential factor governing the structure and
evolution of stars \cite{ClaytonBook},\cite{Carso&Mayer} as well as laboratory
plasmas\cite{Ros92}. In plasmas containing medium to high Z elements, at least
some of the electrons remain bound to the ions even at very high temperatures
and densities (e.g. iron at the center of the sun). As was first recognized by
Edward Teller\cite{Mayeronteller}, in part of the spectral range, the opacity
in these plasmas is dominated by photo-absorption of dipolar radiative
transitions between electronic states of the ions (line absorption).

The present work focuses on line absorption in plasmas in Local Thermodynamic
Equilibrium (LTE). Historically, the development of the theory and
computational approach to this process required a major theoretical
effort(e.g. \cite{Cowanbook}, \cite{Sobelmanspectraandtransition},
\cite{Niki}). The atomic states are evaluated, in all methods, by a
perturbation expansion, using the radial average potential approximation as
the zero order. In this order, the equation for the many-electron states is
reduced to equations for the one-electron states in an average radial
self-consistent potential due to all other electrons. For light elements,
Schroedinger equation is sufficient, for heavier elements, relativistic
treatment is required and the Dirac equation should be solved. The zero order
many-electron states ("configurations") and energies are characterized by the
occupation numbers of degenerate groups of one-electron states ("shells").
Mutual electron-electron interaction removes the degeneracy and splits the
configurational energy. This effect is evaluated as a first order correction.
i.e. as the sum of expectation values, in the zero order states, of energies
due to coulomb interaction between all pairs of electrons. The well known
Slater integrals represent the radial part of these expectation values.

The evaluation of the absorption coefficient requires a compromise between the
needed spectral resolution and the available computer resources (for a
representative list of codes see e.g. \cite{DGillsComparizon},\cite{WorkOPIII}%
,\cite{WorkOPIV}). Existing methods may be classified according to the
resolution of the description of electronic states, and of the contribution of
transitions between them to the absorption coefficient. The most resolved
treatment available is the Detailed Line Accounting (DLA, e.g. \cite{FAC}).
For complex configurations with many states, the number of transitions becomes
prohibitive for numerical calculations. In this case, one may turn to the
Unresolved-Transition-Arrays (UTA) method. In this method the spectral
absorption coefficient due to a transition array consisting of all
single-electron transitions between a specific pair of configurations is
assumed to be of a Gaussian shape. This method is made practical by the
analytical formulae for the three lowest energy-moments (actually
cumulants\cite{ZwanzigBook}) of the UTA spectrum\cite{BauschReview}, in terms
of reduced matrix elements of the dipole operator, Slater integrals and shell
occupation numbers. For heavy ions, the UTA method becomes unpractical due to
the enormous number of possible configurations. The Super-Transition-Array
(STA) method\cite{STA89},\cite{STAconfwidth95},\cite{OregOperator97}
represents a further compromise which allows the evaluation of opacity with
less computational effort at the cost of spectral coarse graining. The coarse
graining is obtained by grouping shells, with adjacent energies, into
supershells, configurations into superconfigurations (SCs) and correspondingly
transition arrays into supertransition arrays. The relative simplicity of the
evaluation of the coarse-grained spectral absorption coefficient is based on
three assumptions (on top of the UTA assumption):

a) \textit{The basic superconfiguration assumption}: All configurations which
form a superconfiguration share the same radial potential with the same set of
one-particle solutions.

b)\textit{The high-temperature approximation}: The spread of the energies of
configurations within a superconfiguation is much smaller than the plasma
temperature. In this limit the Boltzmann factor which determines the relative
probability for a configuration within a superconfiguration may be evaluated
to zero order only, i.e. as the sum of single-electron energies in the mean
potential. Electron-electron interaction energy adds a superconfigurational
average factor common to all configurations within a superconfiguration. This
corresponds to the use of the Gibbs-Bogoliubov-Feynman bound
\cite{ChandlerBook} as an estimate for the Boltzmann factor.

c) \textit{The unresolved supertransition array assumption}: The spectra of
all UTAs which form a STA merge into a single Gaussian shape.

With these three approximations the summation of contributions of all UTAs to
a STA may be performed analytically \cite{STA89},\cite{STAconfwidth95}. By the
third approximation, one needs only to evaluate the three lowest energy
cumulants of the STA spectrum. By the second approximation, the relative
probabilities of configurations are the same as of those in a system of
independent particles in a potential well \cite{Landsberg}. This enables the
derivation of analytical formulae for the moments (and cumulants) in terms of
Slater integrals and partition functions\cite{STA89},\cite{OregOperator97}
which may be evaluated by recursion relations \cite{STA89}%
,\cite{PrtitionWilson99},\cite{PrtitionWilson2007}.

In reference \cite{BWILSONSTRATONOVICH}, a way to avoid the high-temperature
approximation was shown. An analytical formula was written for the partition
function with the full Hamiltonian including electron-electron interaction.
This result was enabled by the application of the Hubbard-Stratonovich
transformation \cite{Hubbard59}, which eliminates of the quadratic dependence
of the energy on the shell occupation numbers (at the cost of introduction of
an auxiliary random field).

The approximations in assumptions (a)-(c) are controlled by the choice of the
degree of spectral coarse graining. In the extreme choice of one shell in a
supershell, and one configuration in a superconfiguration, and a different
average radial potential with a different set of one-electron states
separately for each configuration, one reaches the UTA limit. Clearly, within
the framework of the UTA model, assumptions (a)-(c) are exactly satisfied. The
opposite extreme choice is of one supershell consisting of all shells and one
superconfiguration consisting of all configurations in all degrees of
ionization. In reference \cite{STA89} it was shown that this choice of maximum
spectral coarse graining reproduces the results of the average atom (AA)
model, presented in reference \cite{Shtein&Shalitin&Ron85}.

As mentioned above, for heavy ions, calculations with the UTA resolution are
impractical due to the enormous number of possible configurations. On the
other hand, the AA model is too crude since it wipes out the spectral
structure observed in experiments \cite{STA89}. The STA method allows for a
tune up of the resolution by an iterative refinement procedure starting from
the AA model, increasing the number of superconfigurations and number of
different radial potentials with different sets of one-particle solutions. The
refinement process is stopped when both the values Rosseland and Planck mean
free path (MFP)\cite{ClaytonBook} converge to constant values. A typical STA
run reaches convergence with a few to a few tens of SCF solutions per degree
of ionization, and many more superconfigurations sharing the same potential
and set of single particle energies and orbitals. Actually, as will be
demonstrated by the numerical example in the present work, when a strict
convergence test based on the spectral details rather than the integrated
values of Rosseland and Planck MFP is imposed, convergence of the STA
refinement process in heavy elements is not fully reached even with half a
million STAs.

In the present work, we derive a formula for the spectrum of
Configurationally-Resolved-Super-Ttransition-Arrays (CRSTA) which represents
the \emph{exact sum} of the spectra of all UTAs constituting the STA and
sharing the same SCF solution. Out of the three assumptions ((a)-(c) mentioned
above), required by the traditional STA method, only the basic
superconfiguration assumption (a) was used in the derivation of our new CRSTA
method. As a consistency check we show that the radiation intensity, average
energy and variance of the standard STAs are recovered from the CRSTA by an
approximation based on a cumulant expansion, truncated at the third term.
i.e., the spectrum of a STA evaluated by the traditional method is the
coarse-grained Gaussian approximation of the spectrum of the corresponding CRSTA.

The plan of the manuscript is as follows: In section \ref{abs}, some well
known formulae required for the evaluation of the absorption coefficient in
terms of the two-time dipole autocorrelation function\cite{Kubo1957}%
,\cite{ZwanzigBook},\cite{ZubarevBook},\cite{Griembook97} expanded in the
eigenstates of the atomic Hamiltonian are summarized. The STA order of
summation is briefly reviewed in section \ref{sum}. In section \ref{RSTA} the
formula for the CRSTA spectrum is derived. This formula is limited to the
simple case where the Boltzmann factor is evaluated only with zero-order
energies and the widths of the UTA are neglected (as in the original STA
paper\cite{STA89}). The derivation of the formula for the general case with
the effect of electron-electron interaction in the Boltzmann factor and with
the inclusion of the width of the UTA is derived in the Appendix. Section
\ref{num} contains a brief description of our new code, for the evaluation of
the spectral absorption coefficient by both the standard STA and the new CRSTA
methods, and a numerical example demonstrating the simplicity efficiency and
accuracy of the new CRSTA method. A Summary and discussion are presented in
section \ref{sumdis}.

\section{The absorption coefficient\label{abs}}

The absorption coefficient, in hot dense plasmas in LTE, may be written in
terms of the two-time autocorrelation function of the atomic many-electron
dipole \cite{ZwanzigBook},\cite{Kubo1957},\cite{ZubarevBook}%
,\cite{Griembook97},\cite{Kubofluctdiss}:%

\begin{equation}%
\begin{array}
[c]{c}%
\mu_{\text{at}}\left(  E\right)  =E^{2}\frac{4\pi}{3}\frac{e^{2}}{\hbar
c}n_{0}\operatorname{Re}\left\{  \int_{0}^{\infty}C_{\text{K}}\left(
\tau\right)  e^{iE\tau/\hbar}d\tau\right\}  .
\end{array}
\label{STAUTAhub2}%
\end{equation}
$\mu_{\text{at}}$ is defined as the fraction of the net absorbed radiation
energy at energy $E$ per unit radiation propagation length. $n_{0}$ is the
atoms number density, $C_{K}\left(  \tau\right)  $ is the two-time
autocorrelation function of the atomic many-electron dipole\cite{Kubo1957}:
\begin{equation}%
\begin{array}
[c]{c}%
C_{\text{K}}\left(  \tau\right)  =\beta trace\left(  \rho_{\text{eq}}d\left(
\tau\right)  \tilde{d}\left(  \beta\right)  \right)  .
\end{array}
\label{STAUTAhub2.2}%
\end{equation}

Eq. (\ref{STAUTAhub2}) is just one of many manifestations of the
fluctuation-dissipation theorem connecting between the response of a given
system to an external disturbance and \ the correlation of internal
fluctuations of the system in the absence of the disturbance
\cite{Kubofluctdiss}. In Eq. (\ref{STAUTAhub2.2}), $\rho_{\text{eq}}$ is the
equilibrium density matrix; $\ d\left(  \tau\right)  $ is the the projection
of the Heisenberg representation of the atomic dipole operator, $\sum
\limits_{i}\vec{r}_{i}$, on the polarization vector of the radiation field,
(averaging over all possible polarizations is implied). $\vec{r}_{i}$ is the
position operator of the $i$'th electron. $\tilde{d}$ denotes the Kubo
transform of $d$:%
\begin{equation}%
\begin{array}
[c]{c}%
\tilde{d}\left(  \beta\right)  =\frac{1}{\beta}\int_{0}^{\beta}d\left(
i\hbar\lambda\right)  d\lambda.
\end{array}
\label{STAUTAhub3}%
\end{equation}

(For the relation between the correlation function $C_{\text{K}}$ and more
commonly used functions such as the symmetrized correlation function $C\left(
\tau\right)  =\frac{1}{i\hbar}trace\left(  \rho_{\text{eq}}\left(  d\left(
\tau\right)  d\left(  0\right)  +d\left(  0\right)  d\left(  \tau\right)
\right)  \right)  $ , see references \cite{Kubo1957},\cite{ZwanzigBook}%
,\cite{ZubarevBook} ).

Expanding the dipole and the density operators in eigenfunctions of the atomic
Hamiltonian, $H$, with energies $\left\{  E_{n}\right\}  $ using:
\begin{equation}%
\begin{array}
[c]{c}%
\left(  \rho_{\text{eq}}\right)  _{nn}=\frac{e^{-\beta\left(  E_{n}%
-Q\mu\right)  }}{%
{\displaystyle\sum\limits_{n}}
e^{-\beta\left(  E_{n}-Q\mu\right)  }},
\end{array}
\label{STAUTAhub4}%
\end{equation}
where $Q$ is the number of electrons, $\mu$ is the chemical potential,
$\beta=1/k_{B}T,$ $T$ is the temperature and $k_{B}$ is the Boltzmann
constant, the formula for $\mu_{\text{at}}$ becomes:%

\begin{equation}%
\begin{array}
[c]{c}%
\mu_{\text{at}}\left(  E\right)  =E^{2}\frac{4\pi}{3}\frac{e^{2}}{\hbar
c}n_{0}\frac{1}{\hbar}\frac{1-e^{-\beta E}}{E}\operatorname{Re}\int
_{0}^{\infty}\vartheta\left(  \tau,\beta\right)  e^{-iE\tau/\hbar}d\tau,
\end{array}
\label{STAUTAhub5}%
\end{equation}

with%
\begin{equation}%
\begin{array}
[c]{c}%
\vartheta\left(  \tau,\beta\right)  =\frac{1}{%
{\displaystyle\sum\limits_{i}}
e^{-\beta\left(  E_{i}-Q\mu\right)  }}%
{\displaystyle\sum\limits_{i,f}}
e^{-\beta\left(  E_{i}-Q\mu\right)  }\left\vert d_{if}\right\vert
^{2}e^{-i\left(  E_{i}-E_{f}\right)  \tau/\hbar}%
\end{array}
\label{STAUTAb6}%
\end{equation}
and $\vec{d}_{if}=\left\langle i|\vec{d}|f\right\rangle $. Eq.
(\ref{STAUTAhub5}) is equivalent to the Fermi golden rule in the form used in
\cite{STA89}\cite{STAconfwidth95}. This can be easily seen by performing the
$\tau$ integral and using the relation $\delta\left(  E_{f}-E_{i}-E\right)
=\frac{1}{\pi}\lim_{\gamma->0}\operatorname{Im}\left\{  \frac{1}{E_{f}%
-E_{i}-E-i\gamma}\right\}  $. However, as will become clear from the results
below, the summation over states becomes an easier task when performed prior
to the $\tau$ integration.

\section{Summation over configurations\label{sum}}

Eq. (\ref{STAUTAb6}) is a formula for $\vartheta$ in terms of the exact
many-electron energies and dipole matrix elements. As described in the
introduction, in practice, these quantities are evaluated by a perturbation
expansion, using the radial average potential approximation as the zero order.
A configuration is a zero order many-electron state described by the
occupation numbers of the shells. Symbolically, a configuration is written as
$C=%
{\displaystyle\prod\limits_{s}}
\left(  n_{s}l_{s}j_{s}\right)  ^{q_{s}}$ where a shell is defined by the
\ principal quantum number $n_{s}$, the orbital angular momentum of the large
component in the Dirac wave function, $l_{s}$, and the total orbital+spin
angular momentum $j_{s}$ ; $q_{s}$ is the occupation number of the shell
(configurations are degenerate states). Mutual electron-electron interaction
lifts the degeneracy and splits the configurational energy. This effect is
evaluated as a first order correction, i.e. as the sum of expectation values,
in the zero order states of energies due to coulomb interaction between all
pairs of electrons. In principle, to evaluate the first-order correction one
should diagonalize the perturbation within each degenerate subspace. The
methods discussed here involve only configurational averaged quantities, i.e.
taking the trace, a process which does not require the diagonalization.

Based on this picture, the STA method \cite{STA89}\cite{STAconfwidth95} splits
the summation in Eq. (\ref{STAUTAb6}) into a few stages. First, the spectrum
is split into the different contributions of one-electron transitions, i.e.:%

\begin{equation}%
\begin{array}
[c]{c}%
\vartheta=%
{\displaystyle\sum\limits_{a,b}}
\vartheta^{ab}\text{,}%
\end{array}
\label{STAUTAb6.2}%
\end{equation}
where $\vartheta^{ab}$ is the spectrum due to all possible transitions in
which an electron transits from the shell $n_{a}l_{a}j_{a}$ to another shell
$n_{b}l_{b}j_{b}$ . The summation is over all configurations $C$ in which the
shell $a$ has at least one electron i.e. the occupation number is $q_{a}>0$
and the shell $b$ has at least one hole, i.e. $q_{b}<2j_{b}+1$ . The summation
over all accessible configurations is further partitioned by introducing an
intermediate summation step over superconfigurations $\Xi$\cite{STA89}%
\cite{STAconfwidth95}:%
\begin{equation}%
\begin{array}
[c]{c}%
\vartheta^{ab}=\frac{1}{N}%
{\displaystyle\sum\limits_{\Xi}}
\left(
{\displaystyle\sum\limits_{C\in\Xi}}
g_{C}\exp\left(  -\beta\left(  E_{C}-Q\mu\right)  \right)  \right)
{\displaystyle\sum\limits_{C\in\Xi}}
\frac{g_{C}\exp\left(  -\beta\left(  E_{C}-Q\mu\right)  \right)  }{%
{\displaystyle\sum\limits_{C\in\Xi}}
g_{C}\exp\left(  -\beta\left(  E_{C}-Q\mu\right)  \right)  }\vartheta_{C}%
^{ab},
\end{array}
\label{STAUTAb6.3}%
\end{equation}
where $N=%
{\displaystyle\sum\limits_{\Xi}}
{\displaystyle\sum\limits_{C\in\Xi}}
g_{C}\exp\left(  -\beta\left(  E_{C}-Q\mu\right)  \right)  =%
{\displaystyle\sum\limits_{\text{All}C}}
g_{C}\exp\left(  -\beta\left(  E_{C}-Q\mu\right)  \right)  $.

Each superconfiguration represents a particular distribution of the electrons
between supershells ( a group of energetically adjacent atomic shells).

In Eq.(\ref{STAUTAb6.3}), $\vartheta_{C}^{ab}$\ represents the contributions
from the transition array $C^{ab},$ composed of all transitions from the shell
$n_{a}l_{a}j_{a}$ in the configuration $C$ to the shell $n_{b}l_{b}j_{b}$, and
$g_{C}$ is the zeroth order degeneracy. Following the STA method, we adopt the
assumption of the UTA approach \cite{BauschReview}, that these transitions
merge into an unresolved spectrum of a Gaussian shape (i.e a UTA). Thus, the
three lowest energy moments of the UTA $f_{C}^{ab}$ $E_{C}^{ab}$ $\left(
\Delta E_{C}^{ab}\right)  ^{2}$ are used to construct the spectrum. In the
context of the present work this means
\begin{equation}%
\begin{array}
[c]{c}%
\vartheta_{C}^{ab}\left(  \tau,\beta\right)  =\text{$f$}_{C}^{ab}\exp\left\{
-\frac{1}{2}\left(  \Delta E_{C}^{ab}\right)  ^{2}\tau^{2}+iE_{C}^{ab}%
\tau\right\}
\end{array}
\label{STAUTAb6.6}%
\end{equation}
and:%
\begin{equation}%
\begin{array}
[c]{c}%
\vartheta_{\Xi}^{ab}\equiv\frac{1}{%
{\displaystyle\sum\limits_{C\in\Xi}}
g_{C}e^{-\beta\left(  E_{C}-Q\mu\right)  }}%
{\displaystyle\sum\limits_{C\in\Xi}}
g_{C}\exp\left(  -\beta\left(  E_{C}-Q\mu\right)  \right)  \vartheta_{C}%
^{ab}\left(  \tau,\beta\right) \\
=\frac{1}{%
{\displaystyle\sum\limits_{C\in\Xi}}
g_{C}e^{-\beta\left(  E_{C}-Q\mu\right)  }}%
{\displaystyle\sum\limits_{C\in\Xi}}
\text{$f$}_{C}^{ab}g_{C}\exp\left\{  \Phi_{C}^{ab}\left(  \left\{  q^{_{C}%
}\right\}  ,\beta,\tau^{2},i\tau\right)  \right\}  ,
\end{array}
\label{STAUTAb6.6.1.1}%
\end{equation}
where%
\begin{equation}%
\begin{array}
[c]{c}%
\Phi_{C}^{ab}\left(  \left\{  q^{_{C}}\right\}  ,\beta,\tau^{2},i\tau\right)
\equiv-\beta\left(  E_{C}-Q\mu\right)  -\frac{1}{2}\left(  \Delta E_{C}%
^{ab}\right)  ^{2}\tau^{2}+iE_{C}^{ab}\tau.
\end{array}
\label{STAUTAb6.6.2}%
\end{equation}
The moments of the UTA are represented in the following compact formulae
\cite{STA89}; The strength of a transition is:%
\begin{equation}%
\begin{array}
[c]{c}%
\text{$f$}_{C}^{ab}=q_{a}^{C}\left(  g_{b}-q_{b}^{C}\right)  \left(
\left\langle a\left\vert \left\vert r\right\vert \right\vert b\right\rangle
\right)  ^{2},
\end{array}
\label{STAUTA25}%
\end{equation}

where $\left\langle a\left\vert \left\vert r\right\vert \right\vert
b\right\rangle $ is the reduced matrix element of the dipole. The
configurational average of the energy is:%

\begin{equation}%
\begin{array}
[c]{c}%
E_{C}=%
{\displaystyle\sum\limits_{s}}
q_{s}\left\langle s\right\rangle +\frac{1}{2}%
{\displaystyle\sum\limits_{s}}
{\displaystyle\sum\limits_{r}}
q_{s}\left(  q_{r}-\delta_{rs}\right)  \left\langle s,r\right\rangle ,
\end{array}
\label{STAUTA26}%
\end{equation}

with%
\begin{equation}%
\begin{array}
[c]{c}%
\left\langle s\right\rangle \equiv\varepsilon_{s}+\left\langle s\left\vert
-V\left(  r\right)  -\frac{Z}{r}\right\vert s\right\rangle
\end{array}
\label{STAUTA27}%
\end{equation}
and%

\begin{equation}%
\begin{array}
[c]{c}%
\left\langle s,r\right\rangle =F^{0}\left(  s,r\right)  -\frac{1}{2}%
\frac{g_{s}}{g_{s}-\delta_{s,r}}%
{\displaystyle\sum\limits_{k}}
\left(  1-\delta_{sr}\delta_{k0}\right)  \left(
\begin{array}
[c]{ccc}%
j_{s} & k & j_{r}\\
1/2 & 0 & -1/2
\end{array}
\right)  ^{2}G^{\left(  k\right)  }\left(  s,r\right)  ,
\end{array}
\label{STAUTA28}%
\end{equation}
where $F^{\left(  k\right)  },G^{\left(  k\right)  }$ are the Slater integrals
corresponding to direct and exchange interaction, $\left(
\begin{array}
[c]{ccc}%
j_{s} & k & j_{r}\\
1/2 & 0 & -1/2
\end{array}
\right)  $ is the $3j$ Symbol, and $l_{s},k,l_{r}$ obey the triangle inequality.

The center of gravity of the UTA is:%

\begin{equation}%
\begin{array}
[c]{c}%
E_{C}^{ab}=D_{0}^{ab}+%
{\displaystyle\sum\limits_{s}}
\left(  q_{s}-\delta_{sa}\right)  \left(  D_{s}^{ab}+\left(  \frac{\delta
_{sa}}{g_{s}-1}-\frac{\delta_{sb}}{g_{s}-1}\right)  \varphi\left(  a,b\right)
\right)  ,
\end{array}
\label{STAUTA29}%
\end{equation}
where:%
\begin{equation}%
\begin{array}
[c]{c}%
D_{0}^{ab}=\left\langle b\right\rangle -\left\langle a\right\rangle ,
\end{array}
\label{STAUTA30}%
\end{equation}%
\begin{equation}%
\begin{array}
[c]{c}%
D_{s}^{ab}=\left(  \left\langle s,b\right\rangle -\left\langle
s,b\right\rangle \right)
\end{array}
\label{STAUTA31}%
\end{equation}
and
\begin{equation}%
\begin{array}
[c]{c}%
\varphi\left(  a,b\right)  \equiv-\sum\limits_{\substack{k\neq0\\\text{even}%
}}g_{a}g_{b}\left\{
\begin{array}
[c]{ccc}%
k & j_{a} & j_{a}\\
1 & j_{b} & j_{b}%
\end{array}
\right\}  \left(
\begin{array}
[c]{ccc}%
j_{a} & k & j_{a}\\
\frac{1}{2} & 0 & -\frac{1}{2}%
\end{array}
\right)  \left(
\begin{array}
[c]{ccc}%
j_{b} & k & j_{b}\\
\frac{1}{2} & 0 & -\frac{1}{2}%
\end{array}
\right)  F^{\left(  k\right)  }\left(  a,b\right) \\
+\sum\limits_{k}\frac{g_{a}g_{b}\delta_{k,1}-3}{3}\left(
\begin{array}
[c]{ccc}%
j_{a} & k & j_{b}\\
\frac{1}{2} & 0 & -\frac{1}{2}%
\end{array}
\right)  ^{2}\frac{1+\left(  -1\right)  ^{l_{a}+l_{b}+k}}{2}G^{\left(
k\right)  }\left(  a,b\right)  .
\end{array}
\label{Bauche85.11a}%
\end{equation}

The variance of the UTA is:%
\begin{equation}%
\begin{array}
[c]{c}%
\left(  \Delta E_{C}^{ab}\right)  ^{2}=%
{\displaystyle\sum\limits_{s}}
\left(  q_{s}-\delta_{sa}\right)  \left(  g_{s}-q_{s}-\delta_{sb}\right)
\left(  \Delta^{2}\right)  _{s}^{ab},
\end{array}
\label{STAUTA29.1}%
\end{equation}
where $\left(  \Delta^{2}\right)  _{s}^{ab}$ is independentof the occupation numbers.

These are all the building blocks necessary for the summation in Eq.
(\ref{STAUTAb6.6.1.1}).

\section{Configurationally-Resolved-Super-Transition-Arrays\label{RSTA}}

For the simplicity of presentation we focus on the case in which the width of
the UTA, $\left(  \Delta E_{C}^{ab}\right)  ^{2}$, as well as the
electron-electron interaction terms in the Boltzmann factor, i.e. in $E_{C}$
(but not in $E_{C}^{ab}$), are ignored. The treatment of the general case is
deferred to the Appendix. Ignoring the width of the UTA, as well as the
electron-electron interaction terms in $E_{C}$, Eq.(\ref{STAUTAb6.6.1.1}) is
reduced to:
\begin{equation}%
\begin{array}
[c]{c}%
\vartheta_{\Xi}^{ab}=\frac{1}{%
{\displaystyle\sum\limits_{C\in\Xi}}
g_{C}e^{-\beta\left(  E_{C}-Q\mu\right)  }}%
{\displaystyle\sum\limits_{C\in\Xi}}
g_{C}e^{-\beta\left(  E_{C}-Q\mu\right)  }\text{$f$}_{C}^{ab}e^{iE_{C}%
^{ab}\tau/\hbar},
\end{array}
\label{STAUTA24}%
\end{equation}
where the zeroth order degeneracy is: $g_{C}=%
{\displaystyle\prod\limits_{s\in C}}
\left(
\begin{array}
[c]{c}%
g_{s}\\
q_{s}^{C}%
\end{array}
\right)  $. The Fourier transform yields the STA spectrum. The standard STA
method \cite{STA89} may be obtained by a cumulant expansion
(\cite{ZwanzigBook}) of Eq. (\ref{STAUTA24}) and truncation at the third
cumulant. This may be seen by writing the Taylor series for the factor
$e^{iE_{C}^{ab}\tau/\hbar}$, to obtain the expansion:
\begin{equation}%
\begin{array}
[c]{c}%
\vartheta_{\Xi}^{ab}=\frac{1}{U_{\Xi}}\text{$f$}_{\Xi}^{ab}\sum\limits_{n=0}%
^{\infty}\frac{1}{n!}\left(  i\tau/\hbar\right)  ^{n}\mu_{n}^{_{\Xi^{ab}}},
\end{array}
\label{moment1}%
\end{equation}
where:
\begin{equation}%
\begin{array}
[c]{c}%
\mu_{n}^{_{\Xi^{ab}}}=\frac{1}{\text{$f$}_{\Xi}^{ab}}%
{\displaystyle\sum\limits_{C\in\Xi}}
g_{C}e^{-\beta\left(  E_{C}-Q\mu\right)  }\text{$f$}_{C}^{ab}\left(
E_{C}^{ab}\right)  ^{n}%
\end{array}
\label{moment2}%
\end{equation}
and $f_{\Xi}^{ab}=%
{\displaystyle\sum\limits_{C\in\Xi}}
g_{C}e^{-\beta\left(  E_{C}-Q\mu\right)  }f_{C}^{ab}$ $=\mu_{0}^{_{\Xi^{ab}}}$.

Using the Gaussianity assumption,
\begin{equation}%
\begin{array}
[c]{c}%
\vartheta_{\Xi}^{ab}\left(  \tau,\beta\right)  =I_{\Xi}^{ab}\frac{1}{2\pi}%
\int_{-\infty}^{\infty}\frac{1}{\sqrt{2\pi\left(  \Delta\varepsilon_{\Xi}%
^{ab}\right)  ^{2}}}\exp\left[  -\frac{1}{2}\frac{\left(  E-\varepsilon_{\Xi
}^{ab}\right)  ^{2}}{\left(  \Delta\varepsilon_{\Xi}^{ab}\right)  ^{2}%
}\right]  e^{iE\tau/\hbar}dE\\
=I_{\Xi}^{ab}e^{-i\varepsilon_{\Xi}^{ab}\tau-\frac{1}{2}\left(  \Delta
\varepsilon_{\Xi}^{ab}\right)  ^{2}\tau^{2}}%
\end{array}
\label{GausAs}%
\end{equation}
yields the total radiation intensity, average energy and variance of the STA.
Explicitly,
\begin{equation}%
\begin{array}
[c]{c}%
I_{\Xi}^{ab}=\left[  \vartheta_{\Xi}^{ab}\left(  \tau,\beta\right)  \right]
_{\tau=0}=\frac{1}{U_{\Xi}}\text{$f$}_{\Xi}^{ab},
\end{array}
\label{moment3}%
\end{equation}%
\begin{equation}%
\begin{array}
[c]{c}%
\varepsilon_{\Xi}^{ab}=\left[  \frac{\partial}{\partial\left(  -i\tau
/\hbar\right)  }\ln\left(  \frac{\vartheta_{\Xi}^{ab}\left(  \tau
,\beta\right)  }{I_{\Xi}^{ab}}\right)  \right]  _{\tau=0}=\frac{\mu_{1}%
^{_{\Xi^{ab}}}}{\mu_{0}^{_{\Xi^{ab}}}},
\end{array}
\label{moment4}%
\end{equation}%
\begin{equation}%
\begin{array}
[c]{c}%
\left(  \Delta\varepsilon_{\Xi}^{ab}\right)  ^{2}=\left[  \frac{\partial^{2}%
}{\partial\left(  -i\tau/\hbar\right)  ^{2}}\ln\left(  \frac{\vartheta_{\Xi
}^{ab}\left(  \tau,\beta\right)  }{I_{\Xi}^{ab}}\right)  \right]  _{\tau
=0}=\frac{\mu_{2}^{_{\Xi^{ab}}}}{\mu_{0}^{_{\Xi^{ab}}}}-\left(  \frac{\mu
_{1}^{_{\Xi^{ab}}}}{\mu_{0}^{_{\Xi^{ab}}}}\right)  ^{2}.
\end{array}
\label{moment5}%
\end{equation}
These results coincide with Eqs. (20),(21),(22) in reference \cite{STA89}. In
the practical application of the standard STA method, one is enforced to
represent the spectrum by a large number of narrow Gaussian STAs, in order to
minimize the error due to the truncation of the series. Typically, a STA run
uses only a few self-consistent potentials (and sets of one-particle states
and energies) for every degree of ionization, but a multitude of STAs. Thus, a
multitude of STAs (and many more UTAs) share the same set of one-particle
states and energies.

Our new CRSTA method avoids the approximation of Gaussian STAs (i.e. the
approximation in the truncation at the second cumulant). This is done by the
application of the mathematical machinery of partition functions of
independent particles directly to the contribution of transition arrays
(actually their Fourier transform), to obtain the exact sum of all UTAs
sharing the same one-particle states and energies. Explicitly, this is done by
absorbing the time dependent exponent in the Boltzmann factor and using
Eqs.(\ref{STAUTA26}), (\ref{STAUTA29}) to write the exponential factors in Eq.
(\ref{STAUTA24}) as:%

\begin{equation}%
\begin{array}
[c]{c}%
g_{C}e^{-\beta\left(  E_{C}-Q\mu\right)  }e^{iE_{C}^{ab}\tau/\hbar}\\
=e^{-i\left(  D_{0}^{ab}-D_{a}^{ab}\right)  \tau/\hbar}%
{\displaystyle\prod\limits_{s\in C}}
\left(
\begin{array}
[c]{c}%
g_{s}\\
q_{s}^{C}%
\end{array}
\right)  e^{-q_{s}^{C}\left\{  \beta\left(  \varepsilon_{s}-\mu\right)
+iD_{a}^{ab}\tau/\hbar\right\}  }\\
=e^{-i\left(  D_{0}^{ab}-D_{\alpha}^{ab}\right)  \tau/\hbar}%
{\displaystyle\prod\limits_{s\in C}}
\left(
\begin{array}
[c]{c}%
g_{s}\\
q_{s}^{C}%
\end{array}
\right)  \left(  X_{s}^{ab}\left(  \beta,\tau\right)  \right)  ^{q_{s}^{C}},
\end{array}
\label{STAUTA32}%
\end{equation}
where:%
\begin{equation}%
\begin{array}
[c]{c}%
X_{s}^{ab}\left(  \beta,\tau\right)  =e^{-\left\{  \beta\left(  \varepsilon
_{s}-\mu\right)  +iD_{s}^{ab}\tau/\hbar\right\}  }.
\end{array}
\label{STAUTA33}%
\end{equation}

Next, we define the "superconfigurational degeneracy vector" with components
$\left(  \vec{g}\right)  _{s}=g_{s}$ and \ the supertransitional "Complex
Pseudo Partition Function" (CPPF):%

\begin{equation}%
\begin{array}
[c]{c}%
U_{\Xi}^{ab}\left(  \vec{g},\beta,\tau\right)  =%
{\displaystyle\sum\limits_{C\in\Xi}}
{\displaystyle\prod\limits_{s\in C}}
\left(
\begin{array}
[c]{c}%
g_{s}\\
q_{s}^{C}%
\end{array}
\right)  \left(  X_{s}^{ab}\left(  \beta,\tau\right)  \right)  ^{q_{s}^{C}}%
\end{array}
\label{STAUTA34}%
\end{equation}

and%

\begin{equation}%
\begin{array}
[c]{c}%
U^{ab}\left(  \vec{g},\beta,\tau\right)  =%
{\displaystyle\sum\limits_{\Xi}}
U_{\Xi}^{ab}\left(  \vec{g},\beta,\tau\right)  .
\end{array}
\label{STAUTA35}%
\end{equation}

Using the well known combinatorial relations $q\left(
\begin{array}
[c]{c}%
g\\
q
\end{array}
\right)  =g\left(
\begin{array}
[c]{c}%
g-1\\
q-1
\end{array}
\right)  $ and $\left(  g-q\right)  \left(
\begin{array}
[c]{c}%
g\\
q
\end{array}
\right)  =g\left(
\begin{array}
[c]{c}%
g-1\\
q
\end{array}
\right)  ,$

one gets:%

\begin{equation}%
\begin{array}
[c]{c}%
\vartheta_{\Xi}^{ab}=\frac{\left(  \left\langle a\left\vert \left\vert
r\right\vert \right\vert b\right\rangle \right)  ^{2}e^{-i\left(  D_{0}%
^{ab}-D_{a}^{ab}\right)  \tau/\hbar}}{U_{\Xi}^{ab}|_{\tau=0}}%
{\displaystyle\sum\limits_{C\in\Xi}}
{\displaystyle\prod\limits_{s\in C}}
\left(
\begin{array}
[c]{c}%
g_{s}\\
q_{s}^{C}%
\end{array}
\right)  \left(  X_{s}^{ab}\right)  ^{q_{s}^{C}}q_{a}^{C}\left(  g_{b}%
-q_{b}^{C}\right) \\
=\frac{\left(  \left\langle a\left\vert \left\vert r\right\vert \right\vert
b\right\rangle \right)  ^{2}e^{-i\left(  D_{0}^{ab}-D_{a}^{ab}\right)
\tau/\hbar}}{U_{\Xi}^{ab}|_{\tau=0}}g_{a}g_{b}X_{a}^{ab}%
{\displaystyle\sum\limits_{C\in\Xi}}
\prod\limits_{s\in C}\left(
\begin{array}
[c]{c}%
g_{s}-\delta_{sa}-\delta_{sb}\\
q_{s}^{C}-\delta_{sa}%
\end{array}
\right)  \left(  X_{s}^{ab}\right)  ^{q_{s}^{C}-\delta_{sa}}\\
=\frac{\left(  \left\langle a\left\vert \left\vert r\right\vert \right\vert
b\right\rangle \right)  ^{2}e^{-i\left(  D_{0}^{ab}-D_{a}^{ab}\right)
\tau/\hbar}}{U_{\Xi}^{ab}|_{\tau=0}}g_{a}g_{b}X_{a}^{ab}U_{\Xi,Q-1}%
^{ab}\left(  \vec{g}-\vec{\delta}_{a}-\vec{\delta}_{b},\beta,\tau\right)  .
\end{array}
\label{STAUTA36}%
\end{equation}
In Eq.(\ref{STAUTA36}), the length of the vector $\vec{\delta}_{a}$ equals to
the number of shells, where all components vanish except for the $a$ component
which has the value of $1$.

Note that the algebraic dependence of the formula for the partition
function,Eq. (\ref{STAUTA34}), on the degeneracies, shell occupation numbers
and \thinspace$X$ is the same as of the standard partition function.
Therefore, it obeys the same recursion relations and is accessible to the
efficient evaluation methods \cite{STA89}\cite{STAconfwidth95}%
,\cite{PartitionBlenski},\cite{PrtitionWilson99},\cite{PrtitionWilson2007}.
Unlike references \cite{STA89}\cite{STAconfwidth95},\cite{PartitionBlenski}%
,\cite{PrtitionWilson99},\cite{PrtitionWilson2007} where the partition
function is used for the evaluation of the energy-moments of the spectrum,
Eqs. (\ref{STAUTA36}) when Fourier Transformed with respect to time expresses
the STA spectrum itself as a partition function.

Finally, the formula for the spectral absorption coefficient is obtained by
using Eqs. (\ref{STAUTA36}) and (\ref{STAUTAhub5}):%
\begin{equation}%
\begin{array}
[c]{c}%
\mu_{\text{at}}\left(  E\right)  =\frac{4\pi}{3}\frac{e^{2}}{\hbar c}\frac
{E}{\hbar}n_{0}\left(  1-e^{-\beta E}\right)  \frac{1}{U|_{\tau=0}}%
{\displaystyle\sum\limits_{\Xi,a,b}}
\left(  \left\langle a\left\vert \left\vert r\right\vert \right\vert
b\right\rangle \right)  ^{2}g_{a}g_{b}\\
\times\operatorname{Re}\int_{0}^{\infty}e^{-i\left(  D_{0}^{ab}-D_{a}%
^{ab}\right)  \tau/\hbar}X_{a}^{ab}U_{\Xi,Q-1}^{ab}\left(  \vec{g}-\vec
{\delta}_{a}-\vec{\delta}_{b},\beta,\tau\right)  e^{-iE\tau/\hbar}d\tau.
\end{array}
\label{STAUTA37}%
\end{equation}

\section{Numerical example\label{num}}

We have written, from scratch, a new numerical code called CRSTA. This code
can calculate the spectral absorption by two optional methods, the standard
STA\cite{STA89},\cite{BlenskiGrimaldiPerrot2000} method and our new CRSTA
method. In the STA method, single-particle energies are used to construct the
real partition functions, which in turn are used together with the Slater
integrals to construct the total intensity, average energy and variance of the
STA. The spectrum of each STA is constructed as a Voigt function, which
accounts for STA width as well as Doppler and electron impact effects. The
absorption spectrum is obtained by the summation of contributions from all
STA's. In the CRSTA method, single-particle energies and Slater integrals are
used to construct the CPPF at different times (Eq. (\ref{STAUTA34})), from
which the spectral absorption coefficient is evaluated by Eq.(\ref{STAUTA37}).

The focus of the the ilustrating example was on the replacement of the
traditional STA method of summation over configurations, which is based on the
Gaussian approximation, by the CRSTA method in which the summation is exact
(within the framework of the basic superconfiguration assumption (a)).
Naturally, important technical and physical issues, which are common to the
traditional and the new methods (such as stable and effecient evaluation of
the partition function, and modeling of plasma effects
(\cite{PartitionBlenski}),(\cite{BlenskiGrimaldiPerrot2000})), where out of
the focus of the present work. These issues where treated, in the present
work, by the simplest possible methods.

In both the new and traditinal methods, the relativistic single-particle
radial eigenfunctions and eigenvalues are obtained from a self consistent
solution (SCF) of the radial Dirac equation with the Hartree-Fock-Slater (HFS)
potential in an ion-sphere. In our illustrating example, we use a SCF
procedure, based on a very simple plasma model, that is briefly decribed as
follow. In the first stage, an AA SCF calculation is done, providing the
chemical potential and the average ionized electrons number. The chemical
potential enforces neutrality within the AA's ion-sphere cell, where the free
electrons density is approximated by Thomas-Fermi model. In the second stage,
a single SCF is calculated for each Q state (containing a single supershell
and superconfiguration). For simplicity, we model each Q state as an
ion-sphere that contains Q electrons and embedded in a uniform free electrons
density, $\rho_{0}$. The former is calculated as the AA's ionized electrons
number divided by the AA's cell volume. The neutrality of each Q state's
ion-sphere is obtained by choosing its radius, $r_{Q}$, as: $Q+4/3\rho_{0}\pi
r_{Q}^{3}=Z$.

The STA method requires two refinement loops. In the external refinement loop,
the number of SCs sharing one-particle solutions with the same HFS potential
is decreased. For example, in the AA limit, only one potential is used for all
possible configurations sharing the same potential. In reference \cite{STA89}
a different potential is used for each degree of ionization. Further
refinement and convergence is obtained when few potentials are used for each
degree of ionization. Yet, further spectral resolution is obtained by an
additional internal loop, in which the number of Gaussian STAs sharing the
same HFS solution is increased. In the CRSTA method, the internal loop of
refinement is not necessary, since the exact \ analytical sum of contributions
from all configurations sharing the same potential is represented by a single
CPPF. On the other hand, the CRSTA method requires evaluation of CPPF on a
time grid. We calulate the CPPFs using the fast recursion formulas of
Bar-Shalom \textit{et al}. \cite{STA89}, by substituting the complex quantity
$X_{s}^{ab}\left(  \beta,\tau\right)  $ rather than the standard $X_{s}\left(
\beta\right)  $ ($=X_{s}^{ab}\left(  \beta,0\right)  $). Of course, the
complex factor of $X_{s}^{ab}\left(  \beta,\tau\right)  $ does not affect the
numerical stabilty/instability of these formulas, therefore, one can use the
same alternative stable methods \cite{PartitionBlenski}-
\cite{PartitionGileron}, by substituting $X_{s}\mapsto X_{s}^{ab}\left(
\beta,\tau\right)  $, when numerical instabilities are expected (in our
numerical example, we freeze several lowest energy shells to be fully
occupied, and account small electrons number to be active in several
supershells, therefore, insability is not expected). In our calculation, for
each SCF potential, $V_{j},$ and allowed transition, $a_{j}\rightarrow b_{j}$,
where $a_{j}$ and $b_{j}$ are bound ionic shells we define a time grid. The
resolution and size of the time grid are determined from the zeroth order
transition energies , and an estimation of the expected width, $\Gamma$. Rapid
oscillations in the integrand are elliminated by \ utilizing the fact that the
Fourier transform (FT) in Eq. (\ref{STAUTA37}) equals to the FT of the
function $X_{a}^{ab}(\tau)U_{\Xi,Q-1}^{ab}\left(  \vec{g}-\vec{\delta}%
_{a}-\vec{\delta}_{b},\beta,\tau\right)  ,$ shifted by the energy $D_{0}%
^{ab}-D_{a}^{ab}$. The time resolution, $\Delta t$, is estimated as
$2\pi/\Delta t=n\times\left\vert \varepsilon_{\beta}-\varepsilon_{\alpha
}\right\vert \times w,$ and the time interval as $\left[  0,2m/\Gamma\right]
$, where typically $\ 2<n<5$, $m\geq4$ and $0.1\leq w<0.3$. In order to
incorporate impact broadening effects, we also multiply by the factor
$\exp(-\Gamma/2\tau)$\ prior to the Fourier Transform. Typically, only some
100-1000 CPPF calculations are required for each transition $a_{j}\rightarrow
b_{j},$in order to obtain a UTA like resolution of the spectrum.

Figure (1) shows the b-b absorption spectrum of the $3p_{3/2}\rightarrow
4d_{5/2}$ transition in a Gold plasma at\ temperature of $1keV$ and density of
$0.5gr/cm^{3}$,(see similar example in p. 227 in Ref. \cite{Niki}). Only zero
order energies are used in the evaluation of the partition function and the
widths of the UTAs are ignored \cite{STA89} (See however reference
\cite{STAconfwidth95} and Appendix \ref{hubstrat}). Also, for the sake of
simplicity we have used a single potential for each degree of ionization,
having the ionic states $Q=11..21$ (shells below \textit{3s} shell remain
fully occupied for these Q-states). The time grids contain 2000 grid points
and resolution of $0.15a.u.$ The electron impact parameter, $\Gamma
\approx0.04a.u.$, is calculated from the simple model of section 7.1.2 in Ref.
\cite{Sobelmanspectraandtransition}. For comparison, we have also evaluated
the spectrum by the STA method with increasing number of supershells, and
correspondingly increasing number of superconfigurations and Gaussian STAs.
Figure (2) focuses on the energy range 3420-3460 eV. Clearly, a convergence is
obtained only when a huge number of STAs (half a million) is used. Note that
in the range 3420-3430 eV even half a million of STAs do not fully reveal the
spectral details. In this case, half a million partition functions and half a
million recursion formulas were calculated for the energy and variance of each
STA. In addition, half a million Voigt profiles were calculated for each STA.
On the other hand, in the CRSTA calculation, only$\ 22,000$ CPPF and $11$ FTs
were calculated. In order to check the analytical convergence of the STA
spectrum to the CRSTA spectru, we have also calculated an extreme case of a
UTA spectrum of a low ionic state, $Q=14$, where only four electrons out of
the $14$ remain unfrozen and play an active combinatorial role. For this
check, the STA spectrum is calculated as a sum of Lorentzian profiles, rather
than Gaussians. The CRSTA and STA spectral profiles in this extreme case (not
shown here) completely coincide.%

\begin{center}
\includegraphics[
height=2.9447in,
width=3.8649in
]%
{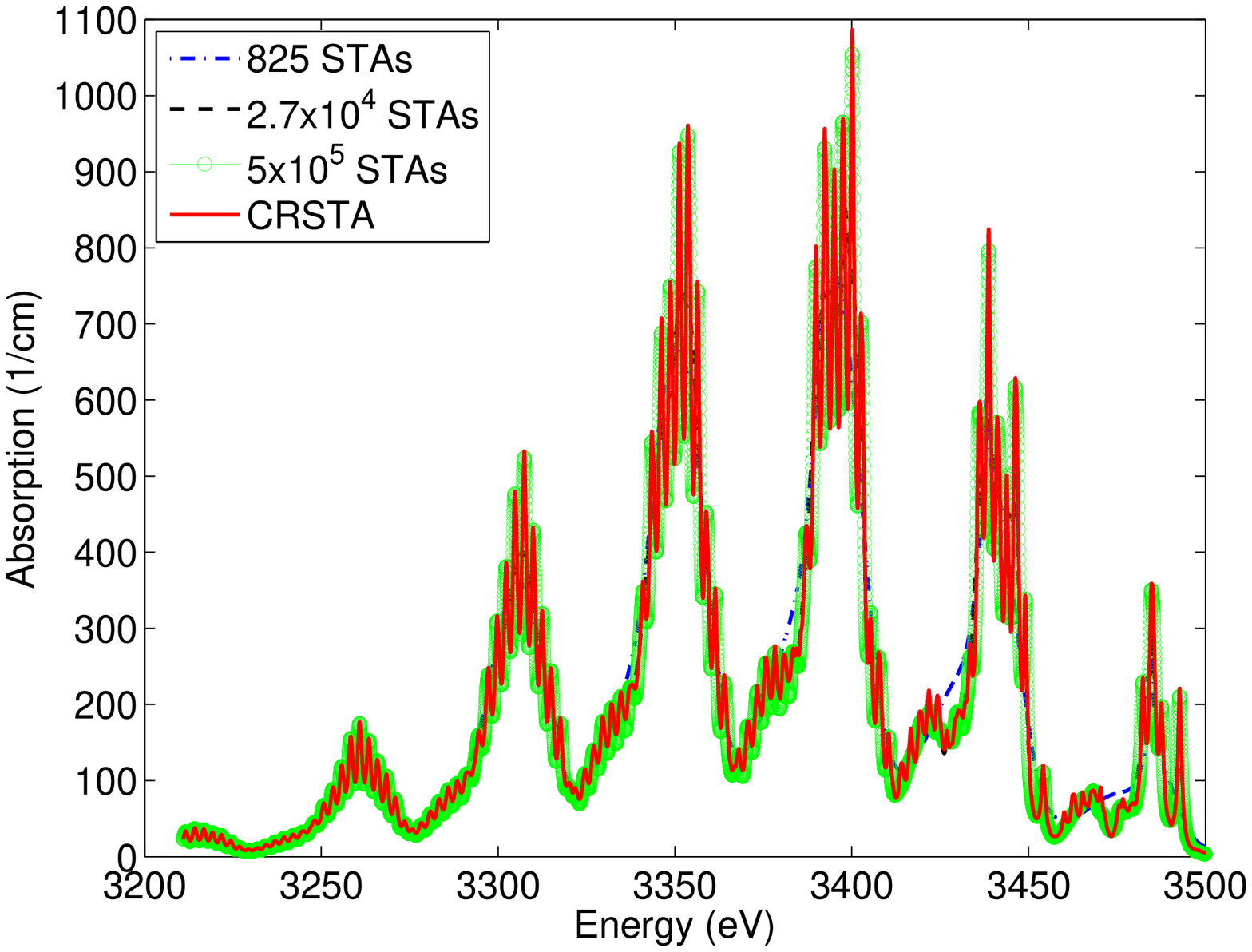}%
\\
Figure(1): b-b absorption spertrum of gold at $T=1keV$ and $\rho=0.5~g/cm^{3}$%
\end{center}

%

\begin{center}
\includegraphics[
height=2.9456in,
width=3.7005in
]%
{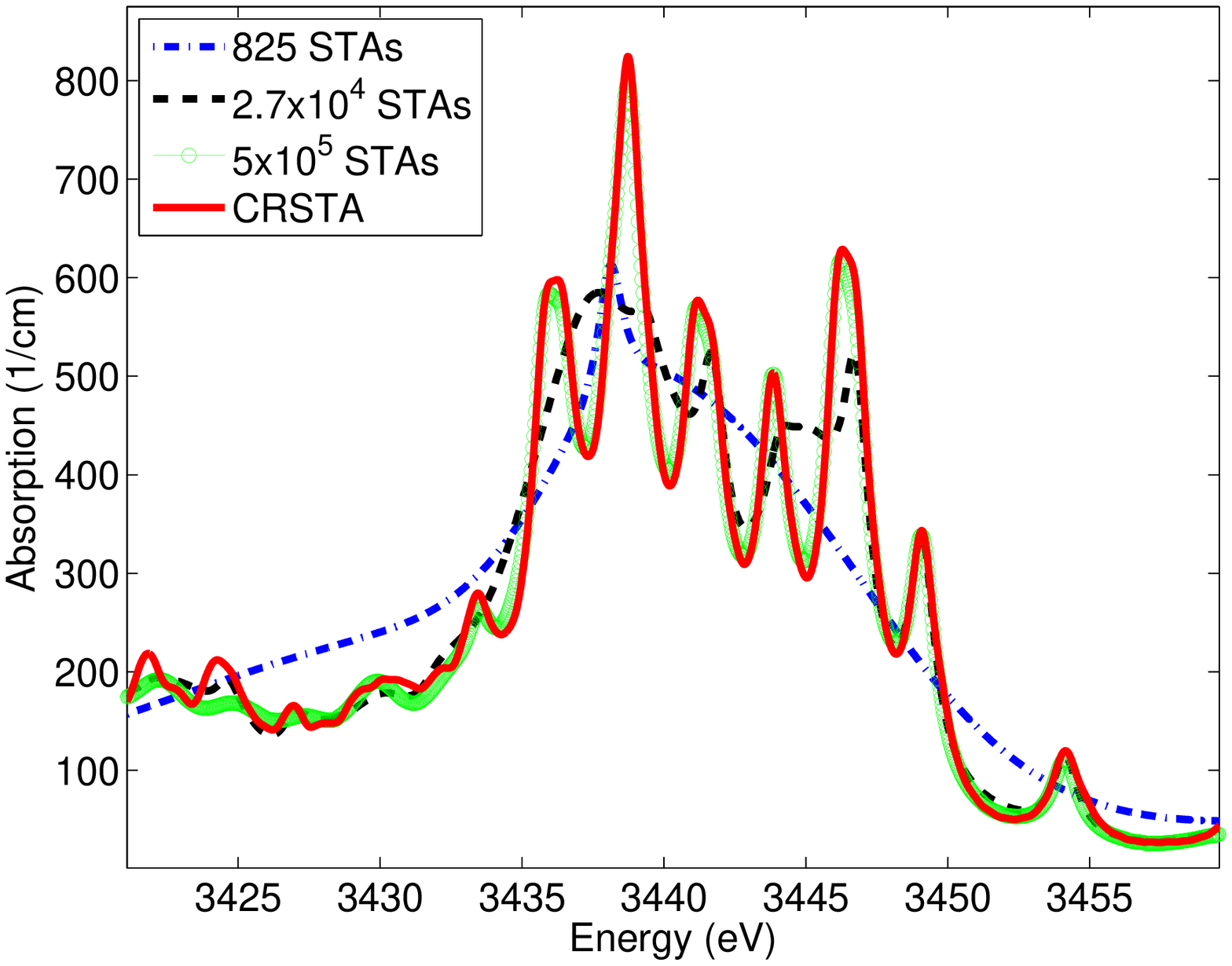}%
\\
Figure(2) Zoom of the energy range 3420-3460 eV in Figure(1)
\end{center}

\section{Summary and discussion\label{sumdis}}

The main result of the present work is in Eqs. (\ref{STAUTA36}) and
(\ref{NN1}),(\ref{NN2}) for the evaluation of the spectrum of a STA in terms
of a single CPPF. In order to analyze the difference between the new CRSTA
method and the traditional STA method let us examine the derivation starting
from Eq. (\ref{STAUTA24}). This equation contains an instruction to sum over
all the transitions constituting the STA, $\Xi$, to get the Fourier transform
of the STA spectrum , $\vartheta_{\Xi}^{ab}\left(  \tau,\beta\right)  $. The
traditional STA formula for $\vartheta_{\Xi}^{ab}$ is obtained by Taylor
expansion with respect to $\tau$, (Eq. (\ref{moment1})). The Gaussian
assumption \ (Eq. (\ref{GausAs})), leads to the truncations of the expansion
at the third term yielding formulae (\ref{moment3}),(\ref{moment4}) and
(\ref{moment5}) for the radiation intensity, average energy and variance from
which the STA spectrum is constructed as a Gaussian. The truncation of the
Taylor expansion at the third term is justified in the short time limit. In
the energy domain, it means a coarse graining of detailed structures finer
than the variance of the STA. In contrast, in the CRSTA method, the summation
to obtain $\vartheta_{\Xi}^{ab}\left(  \tau,\beta\right)  $ is performed
directly by the combinatorial steps (\ref{STAUTA32})-(\ref{STAUTA36}), which
yield the exact result in terms of a single CPPF (Eqs.(\ref{STAUTA36}) and in
the general case equation (\ref{NN1}),(\ref{NN2})). Out of the three
assumptions, required for the derivation of the traditional STA method
(\textit{The basic superconfiguration assumption (a), the high-temperature
approximation (b)} and \textit{the unresolved supertransition array assumption
(c)}), only the first one is used for the derivation of the CRSTA method. i.e.
this new\ method utilizes the simplicity of the analytical manipulations,
enabled by the basic superconfiguration assumption (a), without suffering from
the spectral coarse graining imposed by the unresolved supertransition array
assumption (c). This is the reason why in the numerical example, presented in
figures (1)-(2), a detailed spectrum, which required half a million narrow
STAs for its resolution, is resolved by a few CRSTAs.

The traditional STA concept was originally developed for plasmas at LTE. Later
on, the idea was adapted also to the treatment of non-LTE plasma conditions
(e.g. \cite{SCROLLPHYSREV},\cite{SCROLL}), and also to the treatment of the
electronic degrees of freedom in the equation of state (e.g. \cite{EOSTA}).
The CRSTA method may be adapted also to these tasks.

The possibility to extend the method to resolutions beyond the UTA should also
be explored. Another direction which should be explored is the incorporation
of the CRSTA method with screened hydrogenic model (SHM e.g.\cite{Mayer}) into
a code for rapid (possibly in-line) evaluation of opacity. The CRSTA method
removes the calculational bottleneck of summation over transitions, while the
SHM removes the calculational bottleneck of the SCF process and the evaluation
of Slater integrals.

\begin{acknowledgement}
We thank N. Argaman for a carefull reading of the manuscript and many helpful remarks.
\end{acknowledgement}

\section{Appendix A: A formula for the spectral absorption coefficient
accounting for UTA width and first order correction in the Boltzmann
factor.\label{hubstrat}}

The occupation numbers may be written as a vector of length of the number of
shells, $N_{\text{shell}}$:%
\[%
\begin{array}
[c]{c}%
\left(  \vec{q}^{C}\right)  _{r}\equiv q_{r}^{C}.
\end{array}
\]
Using this definition and the explicit form of the configurational average
energy, center of gravity and variance of a UTA (Eqs.(\ref{STAUTA26}),
(\ref{STAUTA29}) and (\ref{STAUTA29.1}) respectively), the scalar exponent
$\Phi_{C}^{ab}\left(  \left\{  q^{_{C}}\right\}  ,\beta,\tau^{2},i\tau\right)
$ (Eq. (\ref{STAUTAb6.6.2}))\ may be written as a sum of three scalars; a
scalar $\Theta^{ab}$ which is independent of $\vec{q}^{C}$, a scalar product
between $\vec{q}^{C}$ and a vector of coefficients $\vec{\Upsilon}^{ab}$ which
is independent of $\vec{q}^{C}$, and a quadratic form in the vector $\vec
{q}^{C}$ with a real symmetric $N_{\text{shell}}\ast N_{\text{shell}}$ matrix
of coefficients, $\overleftrightarrow{\Omega}^{ab}$, which is independent of
$\vec{q}^{C}$.

The quadratic form prevents a direct application of the combinatorial
manipulations of Eqs (\ref{STAUTA32})-(\ref{STAUTA36}). To cure this problem
we diagonalize the matrix of coefficients, $\overleftrightarrow{\Omega}^{ab}$,
and apply the Hubbard-Stratonovich transformation\cite{Hubbard59} (Eq.
(\ref{HS}) below) which eliminates the nonlinear dependence on occupation
numbers. Explicitly, this is done as follows:

First, the quadratic form is evaluated in a rotated system;
\[%
\begin{array}
[c]{c}%
\vec{p}^{C}=\vec{q}^{C}\cdot\overleftrightarrow{R}^{ab},
\end{array}
\]
which is chosen so that the matrix $\overleftrightarrow{\Omega}^{ab}$ is
diagonal. Explicitly, the elements of the $k$ eigencector of
$\overleftrightarrow{\Omega}^{ab}$ obey:
\[%
\begin{array}
[c]{c}%
\sum\limits_{s}\Omega_{rs}^{ab}R_{sk}=\lambda_{k}R_{rk}.
\end{array}
\]
i.e. $\overleftrightarrow{\Omega}^{ab}$ is diagonalized by a matrix with the
elements $R_{rk},$
\[%
\begin{array}
[c]{c}%
\left(  \overleftrightarrow{\Lambda}\right)  _{jk}\equiv\sum\limits_{s,r}%
R_{jr}\Omega_{rs}R_{sk}=\sum\limits_{r}R_{jr}\lambda_{k}R_{rk}\\
=\lambda_{k}\sum\limits_{r}R_{jr}R_{rk}=\lambda_{k}\delta_{jk},
\end{array}
\]
and $\overleftrightarrow{\Omega}$ is obtained from $\overleftrightarrow
{\Lambda}$ by the inverse of this symmetric transformation:%
\[%
\begin{array}
[c]{c}%
\overleftrightarrow{\Omega}=\overleftrightarrow{R}\cdot\overleftrightarrow
{\Lambda}\cdot\left(  \overleftrightarrow{R}\right)  ^{T}.
\end{array}
\]
Using Eq. (\ref{STAUTA25}) for the dipole matrix element together with the
integral identity:
\begin{equation}%
\begin{array}
[c]{c}%
\exp\left(  -\left(  p_{r}^{C}\right)  ^{2}\lambda_{r}^{ab}\right)
=\sqrt{\frac{c_{r}}{2\pi}}\int_{-\infty}^{\infty}\exp\left(  -\frac{1}%
{2}\left\vert c_{r}\right\vert x^{2}+ixp_{r}^{C}\sqrt{2\lambda_{r}%
^{ab}\left\vert c_{r}\right\vert }\right)  dx
\end{array}
\label{HS}%
\end{equation}
the sum in Eq.(\ref{STAUTAb6.6.1.1}) is written as:
\[%
\begin{array}
[c]{c}%
U_{\Xi}\vartheta_{\Xi}^{ab}=\\
=%
{\displaystyle\sum\limits_{C\in\Xi}}
\left\{  \text{$f$}_{C}^{ab}g_{C}\exp\left\{  -\Theta-\sum\limits_{s}q_{s}%
^{C}\Upsilon_{s}\right\}  \right\}  A_{C}^{ab}%
\end{array}
\]
with%
\begin{equation}%
\begin{array}
[c]{c}%
A_{C}^{ab}=\prod\limits_{r}\exp\left\{  -\left(  p_{r}^{C}\right)  ^{2}%
\lambda_{r}\right\} \\
=\prod\limits_{r}\left\{  \sqrt{\frac{c_{r}}{2\pi}}\int_{-\infty}^{\infty}%
\exp\left(  -\frac{1}{2}\left\vert c_{r}\right\vert x_{r}^{2}+ix_{r}p_{r}%
^{C}\sqrt{2\lambda_{r}^{ab}\left\vert c_{r}\right\vert }\right)
dx_{r}\right\} \\
=\int_{-\infty}^{\infty}...\int_{-\infty}^{\infty}\exp\left(  -\frac{1}{2}%
\sum\limits_{r}\left\vert c_{r}\right\vert x_{r}^{2}+i\sum\limits_{s}q_{s}%
^{C}\sum\limits_{r}x_{r}R_{sr}^{ab}\sqrt{2\lambda_{r}^{ab}\left\vert
c_{r}\right\vert }\right)  \prod\limits_{r}\sqrt{\frac{c_{r}}{2\pi}}dx_{r}\\
=\int_{-\infty}^{\infty}...\int_{-\infty}^{\infty}\exp\left(  -\frac{1}{2}%
\sum\limits_{t}\sum\limits_{s}P_{st}\xi_{s}\xi_{t}+i\sum\limits_{s}q_{s}%
^{C}\xi_{s}\right)  \left\vert \left(  \overleftrightarrow{T}\right)
^{-1}\right\vert \prod\limits_{r}\sqrt{\frac{c_{r}}{2\pi}}d\xi_{r}.
\end{array}
\label{HS1}%
\end{equation}

The matrices $\overleftrightarrow{T}$ and $\overleftrightarrow{P}$ and the
vector $\vec{\xi}$ in Eq.(\ref{HS1}) are defined by:
\[%
\begin{array}
[c]{c}%
\left(  \overleftrightarrow{T}\right)  _{sr}=R_{sr}^{ab}\sqrt{2\lambda
_{r}^{ab}\left\vert c_{r}\right\vert },
\end{array}
\]

\[%
\begin{array}
[c]{c}%
\left(  \overleftrightarrow{P}\right)  _{st}=\left(  \sum\limits_{r}%
c_{r}\left(  \left(  \overleftrightarrow{T}\right)  ^{-1}\right)  _{rs}\left(
\left(  \overleftrightarrow{T}\right)  ^{-1}\right)  _{rt}\right)  ,
\end{array}
\]

and:%
\[%
\begin{array}
[c]{c}%
\vec{\xi}=\overleftrightarrow{T}\cdot\vec{x}.
\end{array}
\]

Next, define:%

\[%
\begin{array}
[c]{c}%
\Gamma^{ab}\left(  \vec{\xi}\right)  =\exp\left(  -\frac{1}{2}\sum
\limits_{t}\sum\limits_{s}P_{st}\xi_{s}\xi_{t}\right)  \left\vert \left(
\overleftrightarrow{T}\right)  ^{-1}\right\vert \prod\limits_{r}\sqrt
{\frac{c_{r}}{2\pi}}.
\end{array}
\]
With this definition, the summation over configuration takes the form:

\begin{equation}%
\begin{array}
[c]{c}%
U_{\Xi}\vartheta_{\Xi}^{ab}=\\
=%
{\displaystyle\sum\limits_{C\in\Xi}}
\left\{  \text{$f$}_{C}^{ab}g_{C}\exp\left\{  -\Theta-\sum\limits_{s}q_{s}%
^{C}\Upsilon_{s}\right\}  \right\}  A_{C}^{ab}\\
=%
{\displaystyle\sum\limits_{C\in\Xi}}
\left\{  \text{$f$}_{C}^{ab}g_{C}\exp\left\{  -\Theta-\sum\limits_{s}q_{s}%
^{C}\Upsilon_{s}\right\}  \right\} \\
\times\int_{-\infty}^{\infty}...\int_{-\infty}^{\infty}\exp\left(  -\frac
{1}{2}\sum\limits_{t}\sum\limits_{s}P_{st}\xi_{s}\xi_{t}+i\sum\limits_{s}%
q_{s}^{C}\xi_{s}\right)  \left\vert \left(  \overleftrightarrow{T}\right)
^{-1}\right\vert \prod\limits_{r}\sqrt{\frac{c_{r}}{2\pi}}d\xi_{r}\\
=\exp\left\{  -\Theta\right\}  \left(  \left\langle a\left\vert \left\vert
r\right\vert \right\vert b\right\rangle \right)  ^{2}\\
\times\int_{-\infty}^{\infty}...\int_{-\infty}^{\infty}\left\{
{\displaystyle\sum\limits_{C\in\Xi}}
q_{a}^{C}\left(  g_{b}-q_{b}^{C}\right)
{\displaystyle\prod\limits_{s\in C}}
\left(
\begin{array}
[c]{c}%
g_{s}\\
q_{s}^{C}%
\end{array}
\right)  \left[  \exp\left\{  -\left(  \Upsilon_{s}-i\xi_{s}\right)  \right\}
\right]  ^{q_{s}^{C}}\right\}  \Gamma^{ab}\left(  \vec{\xi}\right)
d^{N_{\text{shell}}}\vec{\xi}.
\end{array}
\label{NN1}%
\end{equation}

where the expression $g_{C}=%
{\displaystyle\prod\limits_{s\in C}}
\left(
\begin{array}
[c]{c}%
g_{s}\\
q_{s}^{C}%
\end{array}
\right)  $ for the configurational degeneracy was used.

Now define%
\[%
\begin{array}
[c]{c}%
\tilde{X}_{s}^{ab}\left(  \beta,\tau\right)  =\exp\left\{  -\Upsilon_{s}%
^{ab}+i\xi_{s}^{ab}\right\}
\end{array}
\]
and the complex pseudo partition function:%
\begin{equation}%
\begin{array}
[c]{c}%
\tilde{U}_{Q}^{ab}\equiv%
{\displaystyle\sum\limits_{C\in\Xi}}
{\displaystyle\prod\limits_{s\in C}}
\left(
\begin{array}
[c]{c}%
g_{s}\\
q_{s}^{C}%
\end{array}
\right)  \left(  \tilde{X}_{s}^{ab}\right)  ^{q_{s}^{C}}\\
=%
{\displaystyle\sum\limits_{C\left\{  \text{Sum}\left(  q_{s}^{C}\right)
=Q\right\}  }}
{\displaystyle\prod\limits_{s\in C}}
\left(
\begin{array}
[c]{c}%
g_{s}\\
q_{s}^{C}%
\end{array}
\right)  \left(  \tilde{X}_{s}^{ab}\right)  ^{q_{s}^{C}},
\end{array}
\label{NN2}%
\end{equation}
and apply the same combinatorial steps as in Eq.(\ref{STAUTA36}). The result
is a the generalization of Eq. (\ref{STAUTA37}):%

\begin{equation}%
\begin{array}
[c]{c}%
\mu_{\text{at}}\left(  E\right)  =\frac{4\pi}{3}\frac{e^{2}}{\hbar c}\frac
{E}{\hbar}n_{0}\left(  1-e^{-\beta E}\right)  \frac{1}{U|_{\tau=0}}%
{\displaystyle\sum\limits_{\Xi,a,b}}
\left(  \left\langle a\left\vert \left\vert r\right\vert \right\vert
b\right\rangle \right)  ^{2}g_{a}g_{b}\\
\times\operatorname{Re}\int_{0}^{\infty}d\tau\left\{  \exp\left\{
-\Theta^{ab}\left(  \tau\right)  \right\}  \int_{-\infty}^{\infty}d^{N}%
\vec{\xi}\left\{  \tilde{X}_{a}^{ab}\left(  \vec{\xi}\right)  \tilde{U}%
_{\Xi,Q-1}^{ab}\left(  \vec{g}-\vec{\delta}_{a}-\vec{\delta}_{b},\beta
,\tau,\vec{\xi}\right)  \Gamma^{ab}\left(  \vec{\xi}\right)  \right\}
\right\}  .
\end{array}
\label{HS4}%
\end{equation}
Eq.(\ref{HS4}) is a formula for the spectral absorption coefficient which
accounts for the width of the UTA,\ as well as the first order correction in
the Boltzmann factor without the restriction of the high temperature approximation.

\end{document}